\documentclass[12pt]{iopart}
\usepackage{iopams}
\usepackage{epsfig}

\def\bea{\begin{eqnarray}}  \def\eea{\end{eqnarray}}
\def\beq{\begin{equation}}   \def\eeq{\end{equation}}

\def\beeq{\begin{eqnarray}} \def\eeeq{\end{eqnarray}}

\begin{document}

\title[Fluctuations and the clustering
of color sources]{Fluctuations and the clustering
of color sources}

\author{L. Cunqueiro, E. G. Ferreiro and C. Pajares}

\address{
Departamento de F{\'\i}sica de Part{\'\i}culas,
Universidad de Santiago de Compostela, 15782 Santiago de Compostela,
Spain}

\begin{abstract}
We present our results on multiplicity and $p_T$ fluctuations at LHC energies
in the framework of the clustering of color sources.
In this approach, elementary color sources -strings- overlap forming
clusters, 
so the number of effective sources is modified.
We find that the fluctuations are proportional to the 
number of those clusters. 
\end{abstract}

Non-statistical event-by-event fluctuations in relativistic
heavy ion collisions have been proposed as a probe of phase instabilities
near de QCD phase transition.
The transverse momentum and the multiplicity fluctuations
have been measured at SPS
and RHIC energies.
These
fluctuations show a non-monotonic behavior with the centrality of the
collision: they grow as the centrality increases, showing a maximum
at mid centralities, followed by a decrease at larger centralities.
Different mechanisms
have been proposed in order to explain those data. Here, we will apply the clustering of color sources.
In this approach, color strings are
stretched between 
the colliding partons. Those strings act as color sources of particles
which
are
successively
broken by creation of $q {\bar q}$ pairs from the sea.
The color strings correspond to small areas
 in the transverse space filled
with color field created by the
colliding partons. 
If the density of strings increases, they overlap in the transverse 
space, giving rise to a phenomenon of string fusion and 
percolation \cite{Armesto:1996kt}. 
Percolation indicates that the cluster size diverges, 
reaching the size of the system. 
Thus, variations of the initial state can lead to a
transition from
disconnected to connected color clusters. The percolation point signals the 
onset of color deconfinement.

These clusters decay into particles with mean transverse momentum and
mean multiplicity that
depend on the
number of elementary sources that conform each cluster, and the area
occupied by the cluster.
In this approach, the behavior of the $p_T$ \cite{Ferreiro:2003dw}
and multiplicity \cite{Cunqueiro:2005hx}
fluctuations can be
understood as
follows:
at low density, most of the particles are
produced by individual strings with
the same transverse momentum $<p_T>_1$ and the same multiplicity
$<\mu_1>$, so fluctuations are small.
At large density,
above the critical point of percolation, we have only one cluster, so
fluctuations are not expected either.
Just below the percolation critical density, we have
a large number of clusters formed by different number of strings $n$,
with
different size and thus different $<p_T>_n$
and different $<\mu>_n$
so the fluctuations are maximal. 

The variables to measure event-by-event $p_T$ fluctuations are $\phi$ and
$F_{p_T}$, 
that quantify the deviation of the observed fluctuations from
statistically independent particle emission:
\beq
\phi=\sqrt{\frac{<Z^2>}{<\mu>}}-\sqrt{<z^2>}\ ,
\eeq
where $z_i={p_T}_i - <p_T>$ is defined for each particle and
$Z_i=\sum_{j=1}^{N_i} z_j$ is defined for each event,
and
\beq
F_{p_T} = \frac{\omega_{data} - \omega_{random}}{\omega_{random}},\, \ \ \
\omega= \frac{\sqrt{<p_T^2>-<p_T>^2}}{<p_T>}\ .
\eeq
Moreover, 
in order to measure the multiplicity fluctuations, the variance of the multiplicity distribution scaled to the mean value of
the multiplicity has been used.
Its behavior is similar to the one obtained 
for $\Phi (p_T)$, used to quantify the $p_T$-fluctuations,
suggesting that they are related to each other. 
The $\Phi$-measure is independent of the distribution of number 
of particle sources
if the sources are identical and independent from each other. That is,
$\Phi$ should be
independent of the impact parameter if the nucleus-nucleus collision is
a simple superposition of nucleon-nucleon interactions.

\begin{center}
\begin{figure*}
\begin{minipage}[t]{80mm}
\epsfxsize=8.0cm
\epsfysize=5.5cm
\centerline{\epsfbox{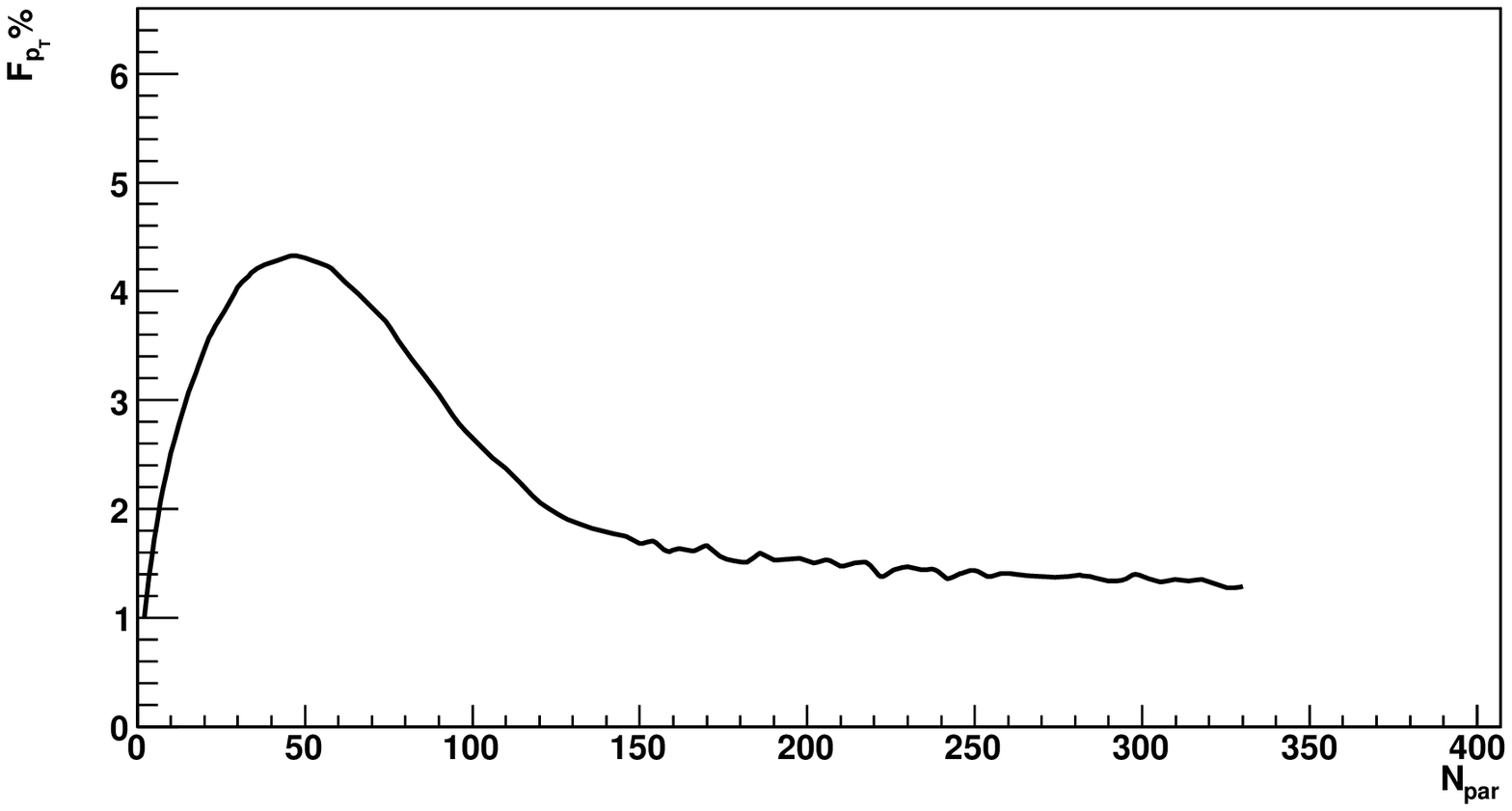}}
\vskip -0.25cm
\caption{$F_{p_T}$ at LHC.}
\end{minipage}
\hspace{\fill}
\begin{minipage}[t]{80mm}
\vskip -5.8cm
\epsfxsize=8.5cm
\epsfysize=6.0cm
\centerline{\epsfbox{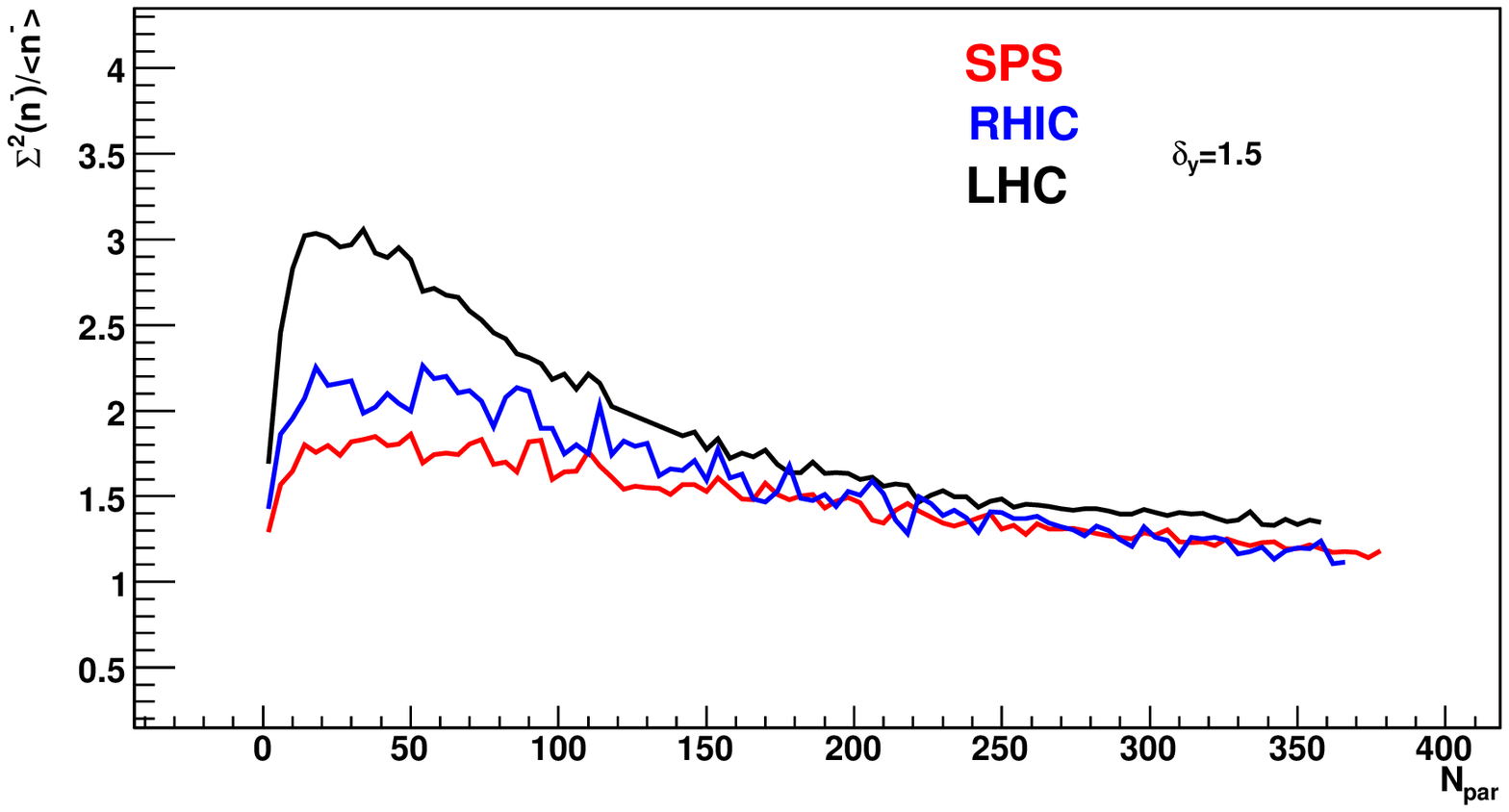}}
\vskip -0.25cm
\caption{Scaled variance on negatively charged particles at, from up to down,
LHC, RHIC and SPS.} 
\end{minipage}
\vskip -0.5cm
\end{figure*}
\end{center}

In Fig. 1 we present our results on $p_T$ fluctuations at LHC. 
Note that the increase of the energy essentially 
shifts the maximum position to a lower number of participants 
\cite{Ferreiro:2003dw}.
In Fig. 2 we show our values for the scaled variance of negatively charged particles
at SPS, RHIC and LHC energies.

Summarizing:
the $p_T$ and multiplicity
fluctuations are due in our approach to
the
different mean $<p_T>$
and mean multiplicities of the clusters,
and they depend essentially on the number
of clusters.
In other words, 
a decrease in the number of effective sources
leads to a decrease of the 
fluctuations.

\end{document}